\documentclass[12pt]{article}

\usepackage{amsmath,amssymb}
\usepackage[margin=1in]{geometry}
\usepackage{setspace}
\usepackage{rotating}
\usepackage{graphicx}
\usepackage{amsfonts}
\usepackage{epsfig}

\begin{document}

\newcommand{\cl}{{\cal L}_{\text m}}
\newcommand{\bea}{\begin{eqnarray}}
\newcommand{\eea}{\end{eqnarray}}
\newcommand{\be}{\begin{equation}}
\newcommand{\ee}{\end{equation}}
\newcommand{\sdr}{\sqrt{a^{\prime}}}
\newcommand{\sdrs}{a^{\prime}}
\newcommand{\xd}{\partial_\tau X}
\newcommand{\xp}{\partial_\sigma X}
\newcommand{\xpp}{\partial_\rho X}
\newcommand{\Xd}{\partial_\tau Y}
\newcommand{\Xp}{\partial_\sigma Y}
\newcommand{\Xpp}{\partial_\rho Y}
\newcommand{\pd}{\dot{\phi}}
\newcommand{\pdd}{\ddot{\phi}}
\newcommand{\cd}{\dot{\chi}}
\newcommand{\cdd}{\ddot{\chi}}
\newcommand{\half}{\frac{1}{2}}
\newcommand{\bi}{\begin{itemize}}
\newcommand{\ei}{\end{itemize}}
\newcommand{\mc}{\mathcal}

\newcommand{\needref}{\textbf{ [REF] }}
\newcommand{\ci}[1]{\left(#1\right)}        
\newcommand{\cu}[1]{\left\{#1\right\}}      
\newcommand{\sq}[1]{\left[#1\right]}        
\newcommand{\an}[1]{\left<#1\right>}        
\newcommand{\bci}[1]{\Bigl(#1\Bigr)}        
\newcommand{\bcu}[1]{\Bigl\{#1\Bigr\}}      
\newcommand{\bsq}[1]{\Bigl[#1\Bigr]}        
\newcommand{\Bci}[1]{\Biggl(#1\Biggr)}      
\newcommand{\Bcu}[1]{\Biggl\{#1\Biggr\}}    
\newcommand{\Bsq}[1]{\Biggl[#1\Biggr]}      
\newcommand{\dbd}[1]{\frac{\partial}{\partial #1}}  
\newcommand{\ef}[1]{\mathrm{d}#1}           
\newcommand{\ket}[1]{\left|#1\right>}       
\newcommand{\bra}[1]{\left<#1\right|}       
\newcommand{\diff}[2]{\frac{{\rm d} #1}{{\rm d} #2}}    
\newcommand{\cm}{\hspace{1cm}}              
\newcommand{\hcm}{\hspace{0.5cm}}           

\newcommand{\cA}{{\cal A}}      \newcommand{\cB}{{\cal B}}
\newcommand{\cC}{{\cal C}}      \newcommand{\cD}{{\cal D}}
\newcommand{\cE}{{\cal E}}      \newcommand{\cF}{{\cal F}}
\newcommand{\cG}{{\cal G}}      \newcommand{\cH}{{\cal H}}
\newcommand{\cI}{{\cal I}}      \newcommand{\cJ}{{\cal J}}
\newcommand{\cK}{{\cal K}}      \newcommand{\cL}{{\cal L}}
\newcommand{\cM}{{\cal M}}      \newcommand{\cN}{{\cal N}}
\newcommand{\cO}{{\cal O}}      \newcommand{\cP}{{\cal P}}
\newcommand{\cQ}{{\cal Q}}      \newcommand{\cR}{{\cal R}}
\newcommand{\cS}{{\cal S}}      \newcommand{\cT}{{\cal T}}
\newcommand{\cU}{{\cal U}}      \newcommand{\cV}{{\cal V}}
\newcommand{\cW}{{\cal W}}      \newcommand{\cX}{{\cal X}}
\newcommand{\cY}{{\cal Y}}      \newcommand{\cZ}{{\cal Z}}

\newcommand{\bA}{{\mathbb A}}   \newcommand{\bN}{{\mathbb N}}
\newcommand{\bB}{{\mathbb B}}   \newcommand{\bO}{{\mathbb O}}
\newcommand{\bC}{{\mathbb C}}   \newcommand{\bP}{{\mathbb P}}
\newcommand{\bD}{{\mathbb D}}   \newcommand{\bQ}{{\mathbb Q}}
\newcommand{\bE}{{\mathbb E}}   \newcommand{\bR}{{\mathbb R}}
\newcommand{\bF}{{\mathbb F}}   \newcommand{\bS}{{\mathbb S}}
\newcommand{\bG}{{\mathbb G}}   \newcommand{\bT}{{\mathbb T}}
\newcommand{\bH}{{\mathbb H}}   \newcommand{\bU}{{\mathbb U}}
\newcommand{\bI}{{\mathbb I}}   \newcommand{\bV}{{\mathbb V}}
\newcommand{\bJ}{{\mathbb J}}   \newcommand{\bW}{{\mathbb W}}
\newcommand{\bK}{{\mathbb K}}   \newcommand{\bX}{{\mathbb X}}
\newcommand{\bL}{{\mathbb L}}   \newcommand{\bY}{{\mathbb Y}}
\newcommand{\bM}{{\mathbb M}}   \newcommand{\bZ}{{\mathbb Z}}

\input{epsf}
\begin{spacing}{1.3}


\long\def\symbolfootnote[#1]#2{\begingroup%
\def\thefootnote{\fnsymbol{footnote}}\footnote[#1]{#2}\endgroup}
{\phantom .}
\begin{center}
\textbf {\large Cosmological Moduli Dynamics} \\
\vspace{1.5em}

Brian Greene,$^1$\symbolfootnote[1]{{\bf email:} {\it greene@phys.columbia.edu}} Simon Judes,$^1$\symbolfootnote[2]{{\bf email:} {\it judes@phys.columbia.edu}} Janna Levin,$^{1,2}$\symbolfootnote[3]{{\bf email:} {\it janna@astro.columbia.edu}}
Scott Watson,$^3$\symbolfootnote[4]{{\bf email:} {\it watsongs@physics.utoronto.ca}} and Amanda Weltman$^1$\symbolfootnote[5]{{\bf email:} {\it weltman@phys.columbia.edu}} \\
\vspace{1.2em}

$^1$Institute for Strings, Cosmology and Astroparticle Physics, Department of Physics,\\ Columbia University, New York, NY. \\
\vspace{0.4em}
$^2$Department of Physics and Astronomy, Barnard College,\\ Columbia University,  New York, NY. \\
\vspace{0.4em}
$^3$Physics Department, University of Toronto,
Toronto, ON. \\
\end{center}
\normalsize
\vspace{1em}

\begin{abstract}
\begin{spacing}{1.5}
Low energy effective actions arising from string theory typically contain many scalar fields, some with a very complicated potential and others with no potential at all. The evolution of these scalars is of great interest. Their late time values have a direct impact on low energy observables, while their early universe dynamics can potentially source inflation or adversely affect big bang nucleosynthesis. Recently, classical and quantum methods for fixing the values of these scalars have been introduced. The purpose of this work is to explore moduli dynamics in light of these stabilization mechanisms.
In particular, we explore a truncated low energy effective action that models the neighborhood of special points (or more generally loci) in moduli space, such as conifold points, where extra massless degrees of freedom arise. We find that the dynamics has a surprisingly rich structure --- including the appearance of chaos --- and we find a viable mechanism for trapping some of the moduli.
\end{spacing}
\end{abstract}

\newpage

\parskip 1ex


\section{Introduction}

One important obstacle to the construction of realistic models of string phenomenology is the generic existence of moduli in the low energy theory \cite{Coughlan:1983ci, Ellis:1986zt, deCarlos:1993jw, Banks:1993en, bbsm, obggr}. The compact manifolds satisfying the string equations of motion generally come in continuous families whose parameters (controlling deformations of the metric and $p$--form fields) become scalar fields with exactly flat potentials in the 4 dimensional effective theory. If on--shell branes are present, there are also moduli that parametrize their relative positions and orientations. Additionally, irrespective of the compact manifold, low energy string models always contain the massless dilaton.

These fields can cause a number of tensions with observation. For example, their presence in the early universe can modify the abundances of hydrogen and helium, a well confirmed prediction of big bang nucleosynthesis \cite{Hashimoto:1998mu}. If there are exactly massless fields at late times, thermal and quantum fluctuations can cause time variation of standard model parameters and 5th force type violations of the equivalence principle \cite{Damour:2001fn, willbook, fischbach}. On the other hand, fields with relatively flat potentials can be useful in building models of slow--roll inflation. For these reasons it is important to explore the variety of moduli space dynamics as fully as possible.

Proposals to mitigate the phenomenological issues mentioned above typically involve either the introduction of a potential for the moduli fields \cite{sethi,Gukov:1999ya, kklt, banks, cham, ob2}, or suppression of their couplings to matter \cite{pol, damnord, sean}. An example of the former strategy that has generated much interest recently is to imagine the universe settling into a vacuum with nontrivial flux threading cycles in the compact extra dimensions \cite{fluxcompactification,sethi}. This induces a potential on the moduli space \cite{sethi,Gukov:1999ya} with a great number of apparently consistent vacuum solutions \cite{Bousso:2000xa, Sethi}: perhaps an infinite number \cite{DeWolfe:2005uu}, or perhaps merely $10^{\sim 300}$ \cite{Ashok:2003gk}\footnote{It is worth keeping in mind that the landscape is not an established feature of string theory. A large number of apparently consistent vacuum solutions have been constructed, but their consistency has not been demonstrated beyond doubt, and there remain worries about whether they should be regarded as separate vacua of a single theory.\cite{Banks:2003es} }. In light of statistical analyses of this `landscape' that find large numbers (but small compared with the total number of vacua) with small positive cosmological constant, some have reconsidered the anthropic framework to explain details of our particular universe \cite{Bousso:2000xa, Susskind:2003kw} that have resisted more traditional approaches.

Moduli space dynamics enter the discussion in two qualitatively different ways.
Some approaches are aimed at avoiding anthropic reasoning through dynamical selection, while others seek to bolster the anthropic approach by providing an underlying mechanism to populate a vast arena of possible universes.
An example of the former strategy is the attempts to find a wavefunction on the moduli space \cite{Ooguri:2005vr,Brustein:2005yn}, whose square might be interpreted as a dynamical weight on the landscape. An example of the latter is the application in \cite{Bousso:2000xa} of results on semiclassical tunneling \cite{bt} to the case of transitions between flux vacua with different cosmological constants.
 
 It was suggested in \cite{Kane:2005cd,Watson:2006px} that instead the problem lies with expanding about a single minimum of the flux potential and that the actual ground state is a relaxed superposition of all of the degenerate connected minima.

The approach we take here is to study moduli dynamics directly in the low energy effective field theory, focusing not on generic loci in moduli space, but on neighborhoods of points where extra massless degrees of freedom appear.
Section \ref{model} motivates our choice of action by considering as an example the effective field theory resulting from a string compactification on a Calabi--Yau 3 fold near a conifold point.
Sections \ref{classical} and \ref{quantum} study two proposed mechanisms for stabilizing the moduli near these so--called \emph{extra species points}, or ESPs. The idea of classical trapping has been considered in a number of contexts; early work in this direction includes \cite{kaloper} while some more recent examples are gases of wrapped strings and branes (see \cite{Battefeld:2005av} and references therein), black hole attractors \cite{kaloper2}, D-brane systems \cite{abel},  M-theory matrix models \cite{Helling:2000kz}, flop transitions \cite{Jarv:2003qx} and conifold transitions \cite{Mohaupt:2004pq,Jarv:2003qx,Mohaupt:2004pr} studied in a cosmological background. In the case of quantum trapping, previous works include \cite{Kofman:2004yc, watson, sera, Silverstein:2003hf, Friess:2004zk}. In \cite{Kofman:2004yc, dimopolous} trapping is used to study trapped inflation and in the latter trapped quintessence. Here, in the conifold context, we go further than previous studies by taking account of both classical and quantum considerations, and of prime importance, establishing that the moduli systems we consider exhibit chaotic dynamics. Chaotic moduli evolution is something that had previously been suggested for moduli dynamics in flat spacetime; we use the Poincar\'{e} `surface of section' technique to put this idea on a firm footing. Surprisingly, it turns out that an understanding of the chaos that appears in the Minkowski case is crucial to interpreting the dynamics when Hubble friction from the gravitational background is included. Specifically, we find the role of Hubble friction in classical moduli trapping to be less effective than previously believed, and that quantum particle production traps the moduli (or at least stabilizes their expectation values) far more efficiently.

To facilitate a direct comparison with refs \cite{Mohaupt:2004pq, Mohaupt:2004pr}, we work with a 5d anisotropic background. Our qualitative conclusions remain the same in the 4d isotropic case.

\section{String Theory Motivation} \label{model}

The low energy effective action we use in the remainder of this paper is relevant to many moduli systems. In this section we outline one context in which it arises --- a Calabi--Yau compactification of M--theory.\footnote{The full derivation contains many technical details of little relevance here, but can be found in \cite{Mohaupt:2004pq,Jarv:2003qx,Mohaupt:2004pr}.} Readers primarily interested in the dynamics of the low energy theory, can skip to the result, which is equation (\ref{lag}).

Our starting point is the low energy description of M-theory, which is the effective field theory of $11$d supergravity with action
\be
S_{11}=\frac{1}{2 \kappa_{11}^2} \left( \int d^{11} x \sqrt{-g_{11}} \left[  \cR_{11} -\frac{1}{48} F_4^2 \right] + \frac{1}{6} \int  A_3 \wedge F_4 \wedge F_4 \right),
\ee
where $F_4$ is the four-form flux $F_4=dA_3$, and $\kappa_{11}^2=8 \pi G_{11}=8 \pi l_p^{9}$, with $G_{11}$ the eleven-dimensional Newton's constant and $l_p$ the Planck length. We want to consider compactification to $5D$ on a Calabi-Yau (CY) 3-fold ${\cal K}$ with Ricci--flat metric $g_{ab}$. Following \cite{Mohaupt:2004pq}, we will take the 11d metric to be of the form,
\begin{align}
    \ef{s^2} = -\ef{t}^2 + a(t)^2\ef{\vec{x}}^2 + b(t)^2\ef{y}^2+ g_{ab}(z)\ef{z^a}\ef{z^b} \quad\quad,
\end{align}
where the spacetime coordinates are $x^\mu=(t,x^1,x^2,x^3,y)$ and the CY coordinates are denoted by $z^a$.

Performing a Kaluza--Klein reduction to 5 dimensions and keeping only the massless modes results in an $N=2$ abelian gauge theory\footnote{Here we are using the language of 4D supersymmetry. In 5 dimensions, $N=2$ is the minimal nontrivial supersymmetry, and so is sometimes referred to as $N=1$. The unambiguous statement is that there are 8 supercharges.} containing; (see e.g. \cite{Cadavid:1995bk})

1 gravity multiplet,
\be
(e_{a\mu},\psi_{\mu I},A_\mu), \;\;\;\;\;\;\;\;\;\;\;\; (\, I=1,2 \, )
\ee
$h_{1,1}-1$ vector multiplets,
\be
(A_\mu^i,\lambda_I^i,\phi^i),     \;\;\;\;\;\;  \;\;\;\;\;\; \left( \, i=1, \ldots, (h_{(1,1)}-1) \, \right)
\ee
and $h_{2,1}+1$ neutral hypermultiplets,
\be
(\zeta^{\bar{i}},A^{\bar{i}}_I)  \;\;\;\;\;\; \;\;\;\;\;\; \;\;\;\;\;\; \left(\, \bar{i}=1, \ldots ,2 (h_{(2,1)}+1) \, \right)
\ee
where $h_{i,j} = \dim H^{i,j}_{\bar{\partial}}(\cK)$ are the Hodge numbers of ${\cal K}$.
The scalar fields in these multiplets parametrize the K\"{a}hler and complex structure moduli spaces of ${\cal K}$ controlling its size and shape. Since the gauge theory is abelian and none of the hypermultiplets are charged, the potential for the scalars is exactly flat, which is just to say that they are moduli.

The above is not the full story, however, as the metric on the hypermultiplet moduli space is geodesically incomplete,\footnote{The same is true of the vector multiplet moduli space, but we won't discuss that case here.} i.e. a classical scalar field can reach the boundary in finite proper time. Originally this was seen as a problem\footnote{See for example the introduction of \cite{Strominger:1995cz}.} because boundary points correspond to singular configurations of ${\cal K}$ that arise when cycles on ${\cal K}$ shrink to zero volume. This is potentially troubling because the geometrical singularity gives rise to various divergent coefficients in the effective action. Type II string theory and M theory have a remarkable way of resolving the issue \cite{Strominger:1995cz}\cite{Greene:1995hu}. Briefly, the singularity in the action can be interpreted as arising from incorrectly integrating out brane wrapping modes that are usually massive (since their mass is proportional to the volume of the cycle they wrap), but become massless at these special points in moduli space where cycles collapse to zero volume. Calculations show that by including these new massless modes in the effective action, the physical singularity is resolved.
It is then natural to seek the explicit form for the low energy effective action near such a geometrical singularity \cite{Mohaupt:2004pq}. The crucial point for our purposes is that the brane states are \emph{charged} under some of the vector multiplets, so the modification required to the 4D theory is to introduce some extra charged hypermultiplets.\footnote{Of course the dimension of the parameter space of the scalars will increase, so the hypermultiplet `moduli space' metric will change as well. This change is necessarily nontrivial, since simply taking a product of two quaternion--K\"{a}hler metrics does not result in another quaternion--K\"{a}hler metric. } It then follows from the extended supersymmetry that several of the scalars acquire potentials. The reason is that the scalars in the charged hypermultiplet must couple to the scalars in the vector multiplet corresponding to the charge they possess.

Unfortunately, little is known about the fully quantum corrected hypermultiplet metric\footnote{An indication of the current status of quantum corrections to hypermultiplet metrics can be found in \cite{Rocek:2005ij}} so it is not yet possible to write explicitly the 5D action corresponding to a compactification of type II string theory on a given Calabi--Yau 3--fold $\cK$. The best we can do at present is to note that $N=2$ supersymmetry implies that the  hypermultiplet metric has holonomy group precisely $Sp(n_H)\cdot Sp(1)$. Spaces admitting such metrics are termed \emph{quaternion--K\"{a}hler}, a somewhat confusing name since as manifolds they are quaternionic, but not in general K\"{a}hler, or even complex. Locally symmetric spaces (i.e. Lie group quotients $G/H$) admitting quaternion--K\"{a}hler metrics were studied by Wolf \cite{wolf}, who classified them into a few (infinite) families. For these examples one can find analytic expressions for the metric, but it is unknown whether any of them correspond to the moduli space that would arise from a genuine Calabi--Yau compactification.

We will now outline the derivation of the full 5d effective theory including the additional massless degrees of freedom.
A more detailed discussion of the construction can be found in \cite{Mohaupt:2004pq} and \cite{Jarv:2003qx}.
There it was shown that one can pick the Wolf space $\frac{U(n_H,2)}{U(n_H)\times U(2)}$, and truncate the action so that only two scalar fields are nonzero; one from a vector multiplet, and one from a charged hypermultiplet. The result is a nonlinear sigma model:
\begin{equation} \label{mohact}
    \mathcal{L}\frac{1}{\sqrt{-g}} = \cR(g)  -\frac{1}{2}G^{(h)}\partial_\mu q \partial^\mu q - \frac{1}{2}G^{(v)}\partial_\mu  r\partial^\mu r - V(r, q) \quad\quad,
\end{equation}
where $\cR$ is the Ricci scalar of the 5d metric $g_{\mu\nu}$, $r, \, q$ are scalar fields parametrizing the vector multiplets and hypermultiplets respectively. The metrics $G^{(h)}$ and $G^{(v)}$ are given by:
\begin{align}
       G^{(h)} = \frac{2}{(1-2q^2)^2}  \quad \quad \quad G^{(v)} = \frac{3(2+r^2)}{(2-3r^2)^2}
\end{align}
and the potential $V(r,q)$ is found to be:
\begin{align} \label{mohpot}
    V(r,q) = (48 \pi)^{2/3} \frac{r^2 q^2}{(1-2q^2)(1-3/2r^2)^{3/2} } \quad\quad.
\end{align}
The field $r$ is one of the K\"{a}hler moduli, and can be thought of (at least when it takes positive values) as measuring the size of some 2--cycle in the compact space $\cK$.\footnote{When $r$ is negative, it can be thought of as the volume of a 2--cycle in a manifold related to $\cK$ by a flop transition.} The point $r=0$ therefore corresponds to a singular Calabi--Yau, and in this case the physics is still sensible because of the M2--brane wrapping states which become massless.

Introducing the geodesic coordinates
\begin{align}
         \phi = \frac{1}{\sqrt{3}}{\rm arctanh} \left(\frac{r(r^2-6)}{(r^2+2)^3/2}\right), \quad \quad \quad \chi = {\rm arctanh}(\sqrt{2q}) ,
\end{align}
the action for the $5$d effective theory becomes

\be  \label{lag}
S_5=\frac{1}{2 \kappa_5^2} \int d^5x \sqrt{-g}\cR(g) - \int d^5x \sqrt{-g}   \left( \frac{1}{2}\partial_\mu \phi \partial^\mu \phi + \frac{1}{2}\partial_\mu  \chi\partial^\mu \chi - V(\phi,\chi)  \right)
\ee
where $\kappa_5^2=\kappa_{11}^2 V_{CY}^{-1}=8 \pi G_5$ with $V_{CY}$ the volume of the extra dimensions, $G_5$ the $5D$ Newton constant,
the $5$d Einstein frame metric is  $g_{\mu \nu}$, and the effective potential is given by
\begin{align}
  V(\phi,\chi) &=  \frac{1}{2}g^2\phi^2\chi^2+ \text{higher order} \label{pot}
\end{align}
with the coupling $g^2 = \frac{2}{3}(48\pi)^{2/3}V_{CY}^{-1} \ll 1$ in Planck units, since $V_{CY} \gg l_p^6$ in order that the SUGRA description is valid.

In \cite{Mohaupt:2004pq}, the model (\ref{mohact}--\ref{mohpot}) was used to demonstrate the possibility of conifold transitions occurring dynamically.\footnote{Dynamics near conifold transition points was also investigated in \cite{palti}.} Such a transition amounts to moving between the two vacuum branches: $r=0$ and $q=0$.\footnote{The existence of two flat directions is not a generic property of singular Calabi--Yaus. Such a situation arises only when there are nontrivial homology relations among the collapsed cycles\cite{Greene:1995hu}. A more typical potential contains terms $\propto\phi^4$, not divisible by $\phi^2\chi^2$.} An important insight gained from the numerical studies of \cite{Mohaupt:2004pq} is that the qualitative behavior of the fields and the scale factor(s) doesn't depend on the precise form of the potential. In particular it was insensitive to terms higher order than $\phi^2\chi^2$.\footnote{Here we mean terms higher order in both $\phi$ and $\chi$, not for example $\phi^2\chi^3$ or $\chi^6$.} We therefore expect that if we stay close to the origin, it is sufficient to study the minimal potential with no higher order terms:
\begin{align} \label{minpot}
    V(\phi,\chi) = \frac{1}{2}g^2\phi^2\chi^2
\end{align}

The property of \eqref{lag} of interest here is that the mass of one field depends on the expectation value of another. This feature is not specific to conifold transitions, but arises whenever new massless states appear in the spectrum.  For instance, if the Heterotic string is compactified on a torus at self--dual radius, a $U(1)\times U(1)$ subgroup of the gauge group is enhanced to $SU(2)\times SU(2)$. Another example is a pair of closely separated D--branes. The ground state of a string stretched between them has a mass proportional to the separation, so as the branes coincide, unexcited strings starting and ending on different branes become massless.  A similar phenomenon occurs in the small instanton phase transition \cite{ovrut}, and as indicated in \cite{Kofman:2004yc} there are many other examples.

Though the points in moduli space where \eqref{minpot} applies are numerous, they are considerably fewer than the number of distinct vacua in the `landscape'. So effects that trap the moduli near points like $\phi=\chi=0$ may help to ameliorate the difficulties associated with extracting predictions from string theory. We now examine some of the special dynamics exhibited by motion in the potential \eqref{minpot}, keeping in mind the problem of understanding how the fields $\phi$ and $\chi$ acquire masses consistent with observations.

\section{Classical Moduli Trapping}  \label{classical}

In this section we will think of \eqref{lag} as a classical scalar field theory, minimally coupled to gravity. The dynamics is governed by a potential and by Hubble friction due to the expansion of the universe.\footnote{\emph{Hubble friction} is a somewhat misleading term. The force it refers to is proportional to the \emph{velocity} of the scalar field, so a better mechanical analogue is air resistance. The distinction is important, since the fields cannot come to rest at a place where the slope of the potential is nonzero, as one might assume from the analogy with friction.}

\begin{figure}[hbtp]
  \vspace{9pt}
  \centerline{\hbox{ \hspace{0.0in}
    \epsfxsize=3.0in
    \epsffile{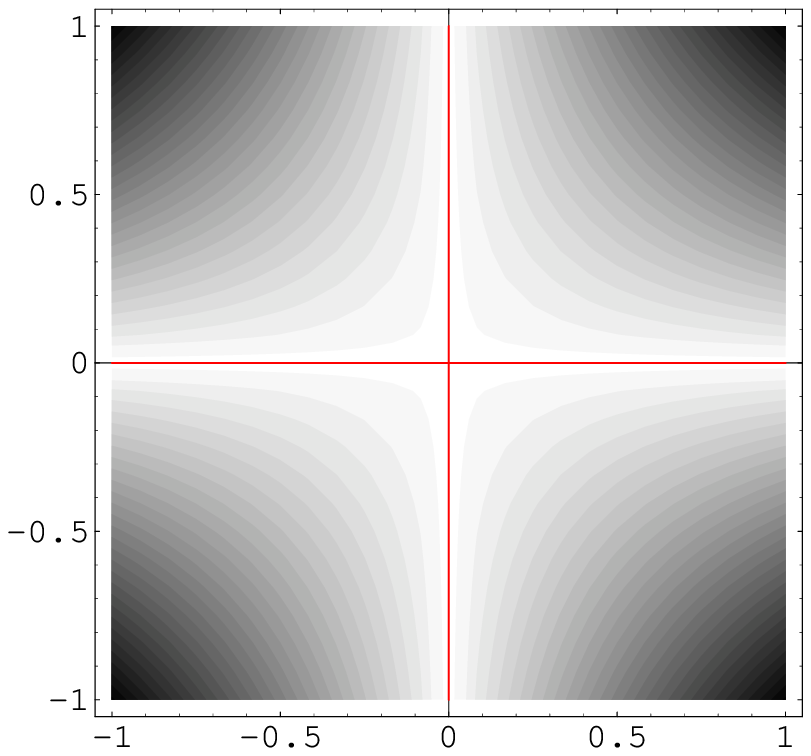}
    \hspace{0.25in}
    \epsfxsize=3.0in
    \epsffile{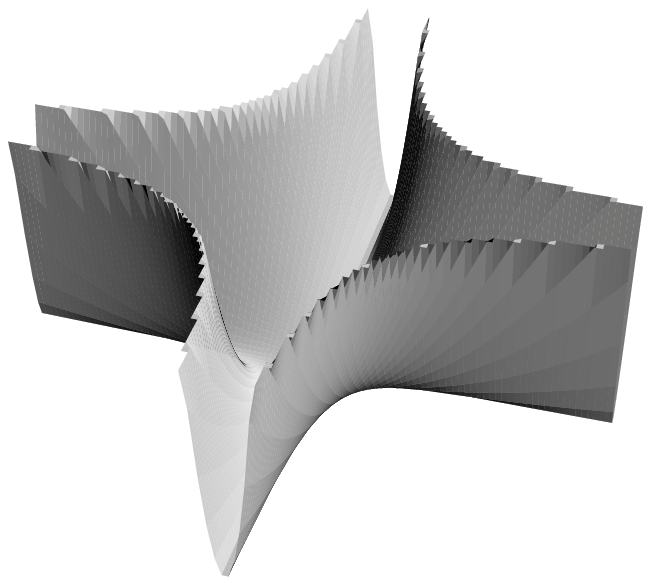}
    }
  }
  \caption{\label{potential} Two views of the potential $V(\phi,\chi)=\frac{1}{2}g^2\phi^2\chi^2$. The vacuum branches (`arms') $\phi=0$ and $\chi=0$, are thehorizontal and vertical axes. Following \cite{Helling:2000kz}, we call the central region the `stadium'. }
\end{figure}

As can be seen from Figure \ref{potential}, the $\phi^2\chi^2$ potential restricts the moduli to a region of their parameter space consisting of two `arms' ($\phi=0$ and $\chi=0$), and a central `stadium'.\footnote{The terminology comes from \cite{Helling:2000kz}.}

The effect of Hubble friction is that in an expanding universe, the scalar fields gradually lose energy, so over time their motion is constrained to smaller and smaller values of $\phi^2\chi^2$. A natural question, then, is whether these two trapping effects, the potential and Hubble friction, are sufficient to mitigate the phenomenological obstacles mentioned in the introduction? This question was studied in \cite{Mohaupt:2004pr}; here we extend the analysis taking particular care of the following two points.

First, a crucial issue is that there is an ambiguity in many discussions as to what is meant by `trapping' and `stabilization'. It is sometimes assumed that all that is required to be consistent with observations is for the motion of the moduli to slow to a stop. In fact, one must also ensure that once they have come to rest, their couplings to matter are very weak, or they have acquired large masses. It is the latter requirement we pursue here. In other words, the eigenvalues of the Hessian of the potential must all be strictly positive (and sufficiently large). For a potential $V=\phi^2\chi^2$, we have:
\begin{align}
    H = \left(\begin{array}{cc}\frac{\partial^2V}{\partial \phi \partial \phi} & \frac{\partial^2V}{\partial \phi \partial \chi} \\\frac{\partial^2V}{\partial \chi \partial \phi} & \frac{\partial^2V}{\partial \chi \partial \chi}\end{array}\right) = \left(\begin{array}{cc}2\chi^2 & 4\phi\chi \\4\phi\chi & 2\phi^2\end{array}\right)
\end{align}
Since $\det H=-12\phi^2\chi^2$ vanishes for vacuum configurations, there is always one eigenvalue identically equal to zero, and so the phenomenological constraints cannot be satisfied. This is easily seen intuitively from Figure \ref{potential}. At any point on either vacuum branch there is a flat direction the field can move in. The best one can do it seems, is to have Hubble friction keep the moduli some way down one of the arms, where only one of them has a mass. We will return to the viability of this option in section \ref{hubblefriction}.

Second, we argue that the classical motion on the moduli space exhibits chaotic dynamics. As we will see, this implies that the classical stabilization found in previous works is only a pseudo-stabilization that traps fields for a finite duration, after which they leave the trapping region and continue their motion.

\subsection{Motion Without Hubble Resistance}

Before taking into account  the effects of a cosmological background, it is instructive to study the dynamics of \eqref{lag} in Minkowski space. In this circumstance $\chi=0$, $\diff{\phi}{t}=\text{const.}$ is a solution of the equations of motion, corresponding to constant velocity motion down one of the vacuum branches. And clearly no matter how long one waits, $\phi$ will not return to the `stadium' area near the origin.  One might guess that the same is true of \emph{almost} straight trajectories where the fields move down one of the arms, but not quite along the minimum of the potential. However, it turns out that any momentum in the $\chi$ direction, no matter how small, is sufficient to return $\phi$ to the origin. This was argued in \cite{Helling:2000kz} as follows. The equations of motion are:
\begin{align} \label{phieqn}
    \ddot{\phi} + (g^2\chi^2)\phi &= 0 \\
    \ddot{\chi} + (g^2\phi^2)\chi &= 0 
\end{align}
where overdots denote derivatives with respect to time.
If $\phi$ is large, and not moving very fast, then $\chi$ is approximately a harmonic oscillator with frequency $g\phi$. Following \cite{Helling:2000kz} we split the energy into two components, one corresponding to motion down the arm, the other to transverse oscillations:
\begin{align}
    E_{\phi} =  \frac{1}{2}\dot{\phi}^2, \cm E_{\chi}= \frac{1}{2}\dot{\chi}^2+\frac{1}{2}g^2\phi^2\chi^2
\end{align}
For any harmonic oscillator one can show that $\left<\chi^2\right>=E_{\chi}/{g^2\phi^2}$, and replacing $\chi^2$ by its expectation value in \eqref{phieqn} results in an effective potential:
\begin{align} \label{effectivepot}
    \ddot{\phi} + \diff{V_{\text{eff}}}{\phi} = 0, \hcm\cm V_{\text{eff}} = E_{\chi}\log(\phi)
\end{align}
For any positive $E_{\chi}$, $V_{\text{eff}}$ increases monotonically, thus returning $\phi$ to the origin.

\subsection{Chaos}

 It has previously been noticed that the potential of Figure \ref{potential} has some of the features associated with chaos \cite{Mohaupt:2004pr}. For example, slightly different initial conditions can result in motion down different arms, and the effective logarithmic potential ensures that generically there is crossing of trajectories. Also since the potential in the stadium region is negatively curved, we can intuitively think of the resulting dynamics as a Newtonian approximation to geodesic motion in a negatively curved space, which is chaotic by the theorem of Horne and Moore \cite{horne}. The importance of these observations for the case of motion with Hubble friction warrants a more concrete analysis of the appearance of chaos\footnote{ Chaotic behavior in cosmologies with scalar fields
that have similar potentials was studied in \cite{levin, barrow}}
, which we now perform using the Poincar\'{e} `surface of sections' method. The result is that indeed we find chaotic scattering in the stadium with almost regular motion in the arms.

\subsection*{Poincar\'{e} Surface of Sections --- Visualizing Chaos}

The Poincar\'{e} surface of sections (henceforth PSS) technique was developed by Birkhoff \cite{Birkhoff} and Poincar\'{e} \cite{Poin}, and has been widely used in studies of chaotic dynamics. We will briefly review the idea of a PSS and then apply the technique to the problem at hand: motion in the potential $V(\phi,\chi)=\frac{1}{2}g^2\phi^2\chi^2$.

\begin{figure}[h]
\begin{center}
\epsfxsize=4.0in
\epsffile{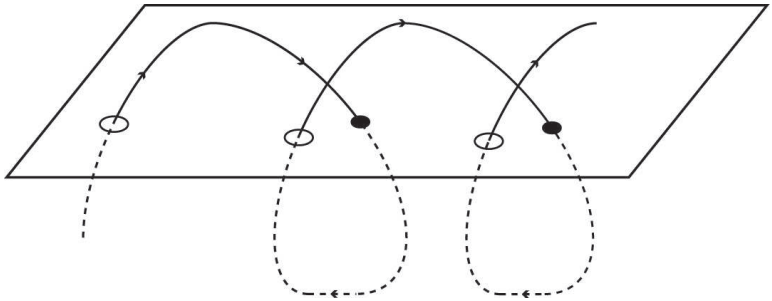}
\caption{\label{PSS} Schematic diagram of a Poincar\'{e} surface of section. The circles show the points where the phase trajectory cuts the hyperplane. Since the hyperplane is oriented, one can distinguish such points by the direction of the phase trajectory when it passes through. A PSS is a collection of such points with the phase trajectory moving in a specified direction. So this diagram really shows part of \emph{two} PSS's --- corresponding to the solid and hollow circles. }
\end{center}
\end{figure}

The PSS is based on the idea that to observe chaotic behavior one need not examine the motion through phase space in all its complexity. Instead one can pick a codimension 1 hyperplane in the phase space and mark each point that's crossed by a phase space trajectory. See Figure \ref{PSS} for a schematic representation of the PSS technique.

It is a well known result of Kolmogorov \cite{Kol}, Arnold \cite{Arnold} and Moser \cite{Moser} known as the KAM theorem that if a Hamiltonian system is integrable, then motion is confined to a torus in phase space. If the motion is bounded, then after a long time (many cuts through the hyperplane), a PSS traces out a planar section of a torus. In our example, the phase space is coordinatized by $\phi,\chi,p_{\phi},p_{\chi}$, and is therefore 4 dimensional. By working on a fixed energy shell we reduce to 3 dimensions by solving for $p_{\chi}$ in terms of the other 3 coordinates and the total energy $E$. A codimension 1 hyperplane is then a 2d space, and if the system is integrable then the PSS is a smooth curve, i.e. a `toric section'.

We choose the hyperplane to be specified by the condition $\chi=0$, and then plot the $\phi,p_{\phi}$ coordinates each time the trajectory pierces the $\chi=0$ slice in the $p_{\chi}>0$ direction. In this way we build up a plot of successive $(\phi,p_{\phi})$ values. A smooth curve indicates the absence of chaos.  If on the other hand the system is not integrable, then we will see an irregular pattern of points.

\begin{figure}[h]
\begin{center}
\epsfxsize=4.0in
\epsffile{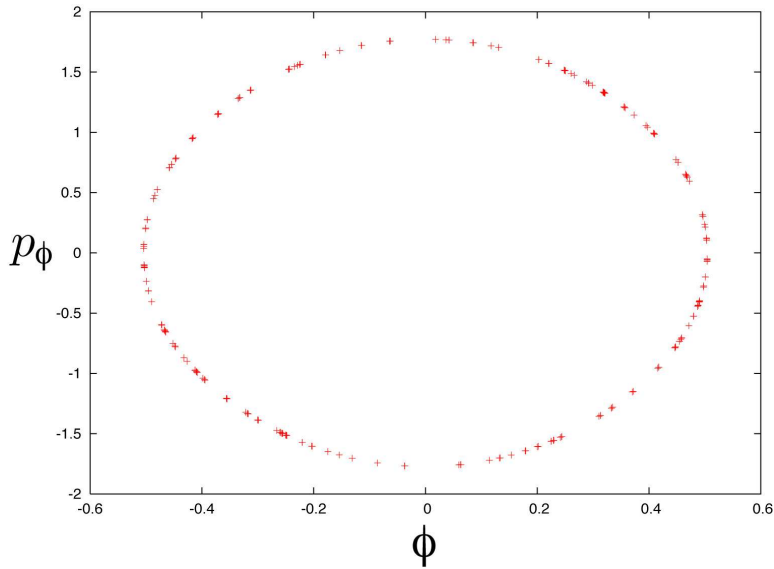}
\caption{\label{psos1} Surface of Sections for $\sigma = 10$, $1/2g^2 = 10$, and $E=20$. The points trace out a smooth curve.}
\end{center}
\end{figure}

\begin{figure}[h]
\begin{center}
\epsfxsize=4.0in
\epsffile{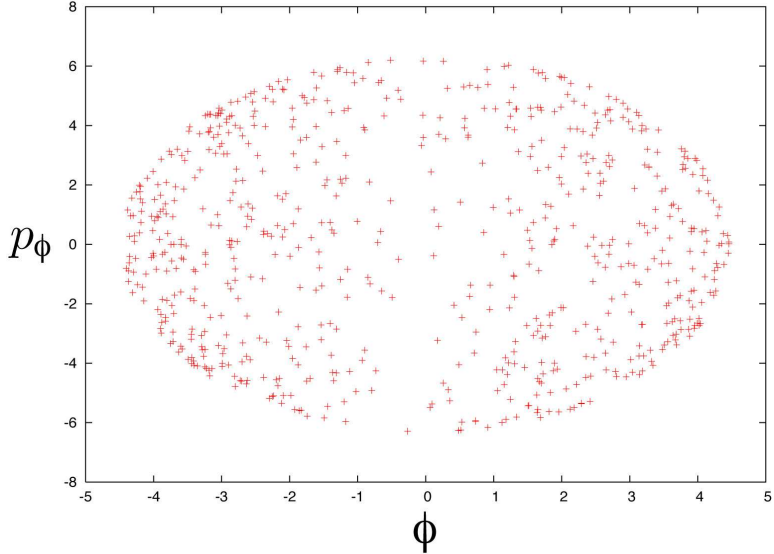}
\caption{\label{psos2} Surface of Sections for $\sigma = 1$, $1/2g^2 = 10$, and $E=20$. Here the points fill out a region of the $\phi,p_{\phi}$ plane, indicating chaotic dynamics. }
\end{center}
\end{figure}

To develop an insight for the onset of chaos we consider the following potential which is related to \eqref{minpot} by the addition of a control term $\sigma(x^2 +y^2)$,
\begin{align}
    V(x,y) = \frac{1}{2} g^2 \phi^2 \chi^2 + \sigma (\phi^2 +\chi^2) \quad\quad.
\end{align}
For $\sigma \gg \frac{1}{2}g^2$ the motion is manifestly not chaotic as the potential is separable. In fact it turns out that $\sigma \sim g^2$ is sufficient to remove any indication of chaos from the PSS. To illustrate this, we will start with $\sigma = \frac{1}{2}g^2 = 10$ and dial $\sigma$ down relative to $g^2$ so that we can see the signature of chaos emerge. We can see this effect clearly by comparing Figure \ref{psos1} to Figure \ref{psos2}, where $\sigma$ has been reduced by a factor of $10$.


\begin{figure}[h]
\begin{center}
\epsfxsize=4.0in
\epsffile{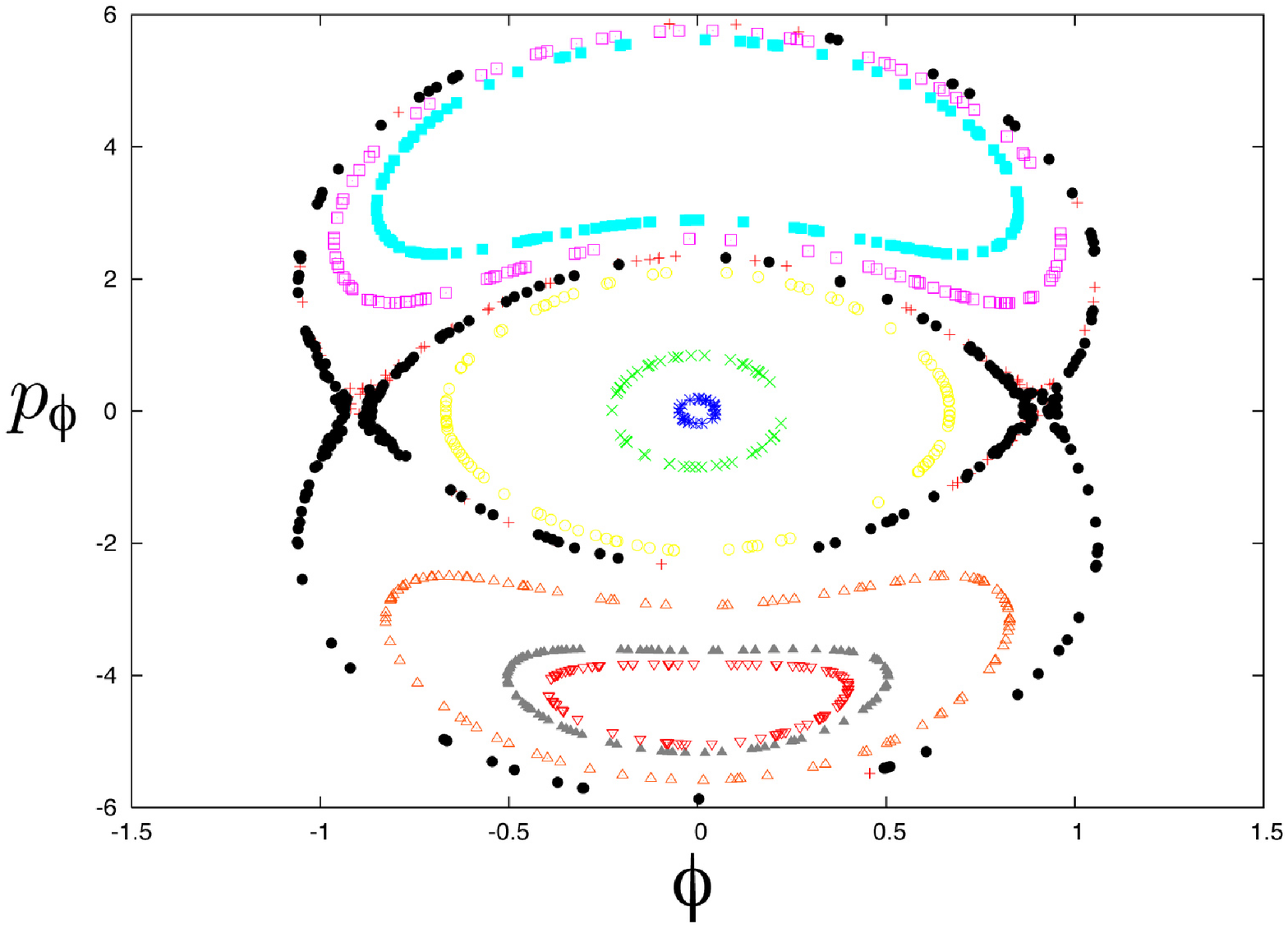}
\caption{\label{psos3} Surface of Sections for  $\sigma =10$, $1/2g^2 = 10$, and $E=20.0$, for different initial conditions.}
\end{center}
\end{figure}

\begin{figure}[h]
\begin{center}
\epsfxsize=4.0in
\epsffile{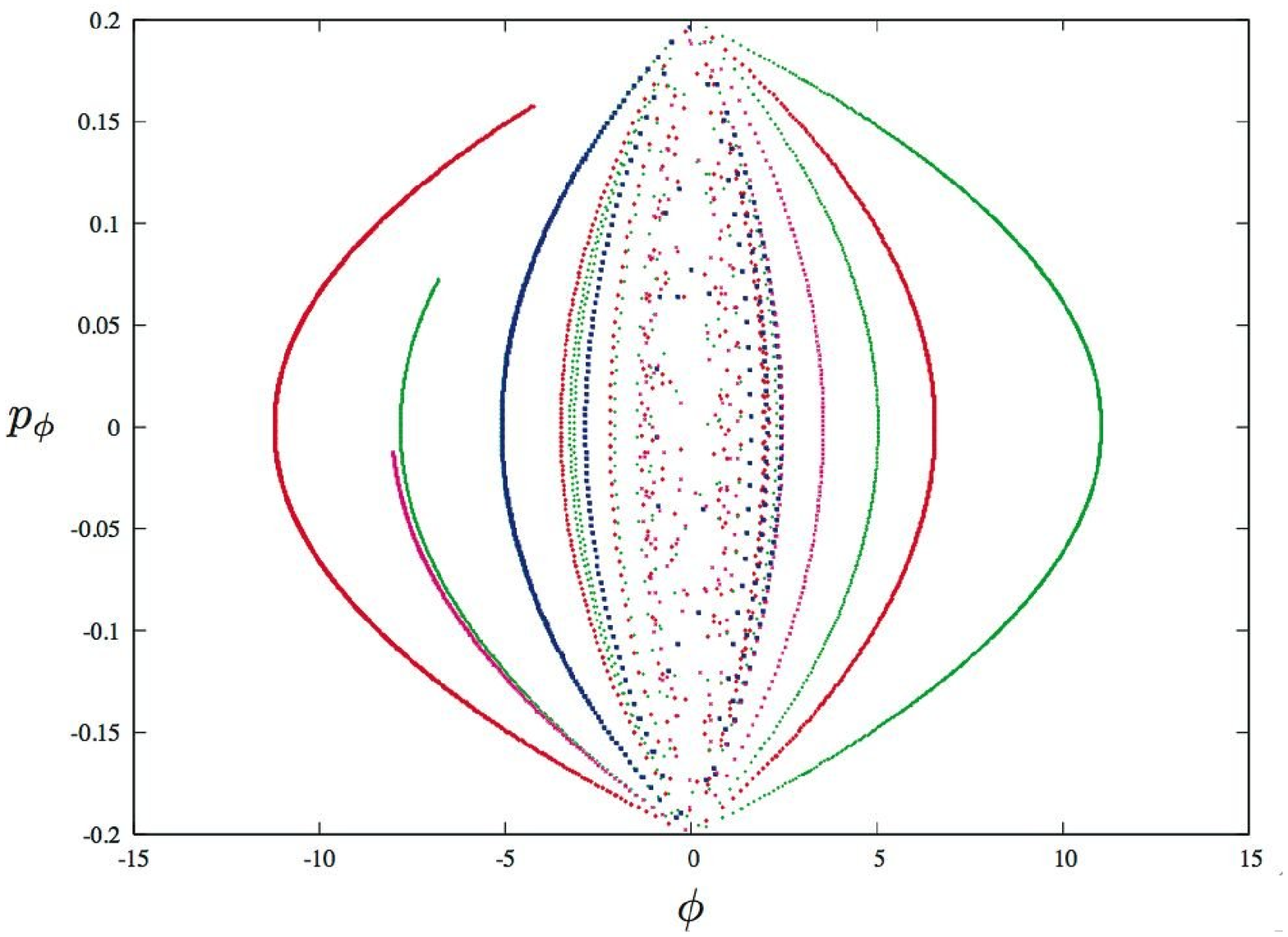}
\caption{\label{psos6} Surface of Sections for  $\sigma = 0$, $g^2 = 20$, and $E=20$ for multiple trajectories.}
\end{center}
\end{figure}

One should note that the smooth curve typical of an integrable system need not be an ellipse as in Figure \ref{psos1}. Figure \ref{psos3} illustrates alternatives.
The case of interest to us where $\sigma$ is exactly zero is more intricate because there are exactly flat directions in the potential. When the motion is far down one of the arms, we have seen that it is governed approximately by a harmonic oscillator in one direction, and a logarithmic potential in the other. We therefore see orbits that are approximately regular for a while, before degenerating into chaotic motion, as seen in Figure \ref{psos6}. The almost regular orbits on the outer shell of the ``onion'' in Figures \ref{psos6} represent the motion up an arm, which is always very close to a regular orbit. The irregular pattern of points filling up the central region indicates the chaotic nature of the motion in the stadium. Thus, though we may find stabilisation in the stadium using further techniques, we cannot use initial conditions to constrain exactly where in the stadium this stabilisation will occur.

The significance of the alternation of chaotic motion with approximately regular orbits is that the trapping mechanism provided just by the potential is actually not effective in stabilizing or trapping any of the moduli. There might be a period of time where the fields are down one of the arms (so that one of them has a mass), but they will always return to the central stadium region before shooting off down another perhaps different arm.

Next we will look at how Hubble friction changes this situation. Since the energy is no longer conserved once we turn on Hubble friction, a PSS analysis is no longer appropriate, however what we've learned in the above section will prove useful in all that follows. Specifically we will find that since the motion in the stadium is chaotic, even once the fields have lost a significant amount of energy to Hubble friction, they will always eventually scatter up an arm and it would thus require extremely special initial conditions to be trapped on a regular orbit.

\subsection{Motion Including Hubble Resistance} \label{hubblefriction}

\begin{figure}[h]
\begin{center}
\epsfig{figure=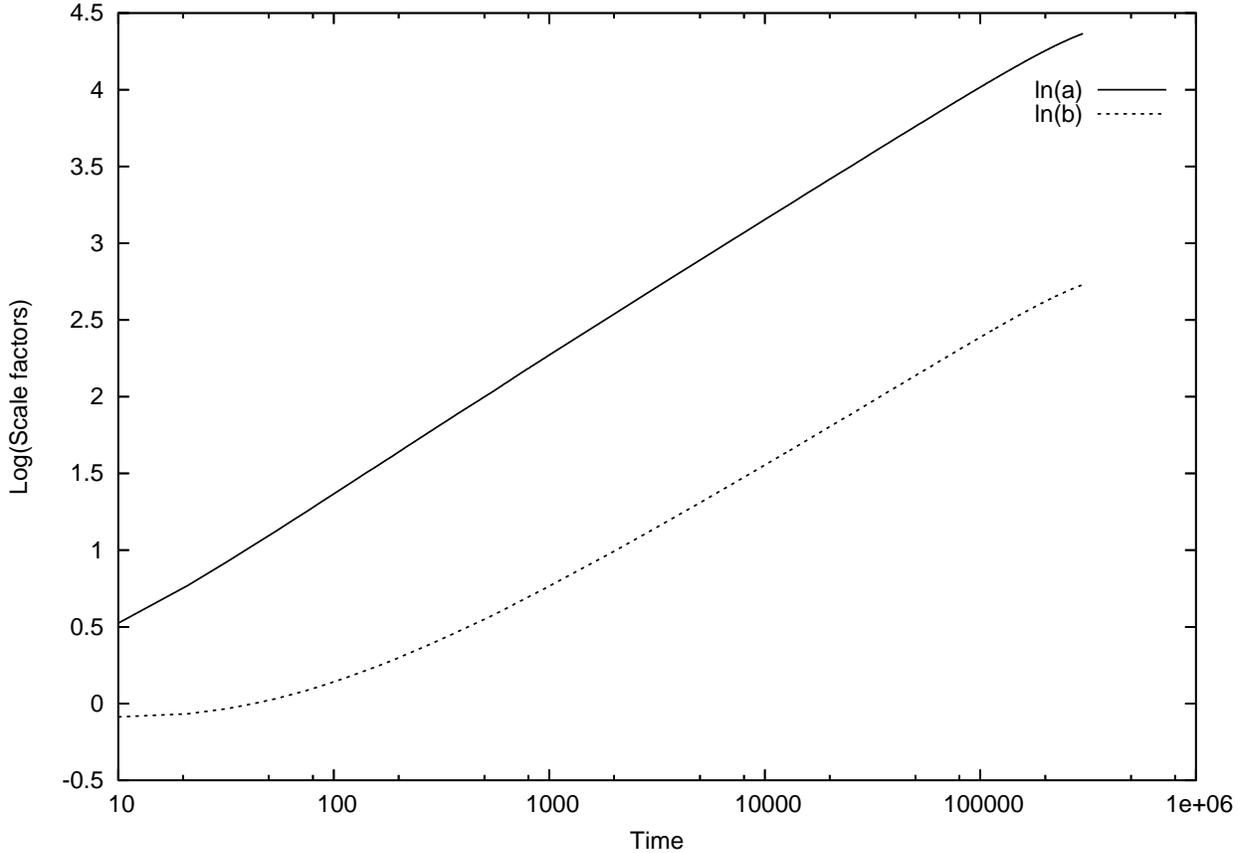,angle=-90,width=\textwidth}
\caption{\label{kas} The scale factors as a function of time. After $t\sim1000$ Planck times the spacetime is well approximated by power law expansion. Initial conditions here are $(\phi,\dot{\phi},\chi,\dot{\chi},a,\dot{a},b)=(0,0.2,0.2,0,1,0.1,1)$ and $\dot{b}$ is determined by the Friedmann equation.}
\end{center}
\end{figure}

In an expanding universe, fields gradually lose energy. One might hope that consequently the fields eventually stop at some point away from the origin, thus leaving one field with a mass. We take the metric to be isotropic in 3 of the space dimensions and homogeneous in all 4:
\begin{equation}
    \ef{s}^2 = -\ef{t}^2 + a(t)^2\ef{\vec{x}^2} + b(t)^2\ef{\vec{y}^2}
\end{equation}
As noted in the introduction, we choose this spacetime so as to compare our results directly with those of \cite{Mohaupt:2004pq,Jarv:2003qx,Mohaupt:2004pr}. Taking the gravitational background into account, the equations of motion become:
\begin{align}
    \ddot\phi + \ci{3H_a+H_b}\dot{\phi} + g^2\chi^2 \phi &= 0 \\ \label{chieqn}
    \ddot\chi + \ci{3H_a+H_b}\dot{\chi} + g^2\phi^2 \chi &= 0 \\
    \dot{H_a} + H_a^2 - H_aH_b +  \frac{2}{3}\kappa_5^2T &=0 \\
    \dot{H_b} + H_b^2 -2H_a^2 + H_aH_b + \frac{2}{3}\kappa_5^2 T &=0
\end{align}
where $T=\frac{1}{2}\ci{\dot{\phi}^2+\dot{\chi}^2}$ is the kinetic energy density of the fields.
After some time, numerical simulations like those shown in Figure \ref{kas} suggest that the spacetime approaches a power law solution:
\begin{align}
    \ef{s}^2 = -\ef{t}^2 + t^{2q}\ef{\vec{x}^2} + t^{2q_4}\ef{y^2}
\end{align}
for some constants $q$ and $q_4$.
The energy density in the fields then satisfies:
\begin{align}
    \rho = 3H_a^2 + 3H_aH_b = \frac{3q(q+q_4)}{t^2}\quad\quad.
\end{align}

We expect that averaged over a sufficient interval, the kinetic energy is some fixed fraction of the total, though not necessarily $1/2$ as for a harmonic oscillator. The velocity of the fields is therefore $\sim \sqrt{\rho} \sim t^{-1}$, and the distance they have travelled $\sim \log(t)$. It follows that the fields move an infinite distance, and do not gradually approach any particular point in the moduli space.  Moreover, since the scattering in the stadium is chaotic, after some time the fields will always eventually end up moving \emph{almost} exactly down an arm, so in this sense Hubble friction does not trap the fields.\footnote{Were the dynamics not chaotic, one can imagine the motion confined to a stable orbit in the stadium which converges to the origin as the energy decreases.}

\begin{figure}[h]
\begin{center}
\epsfxsize=4.0in
\epsfig{figure=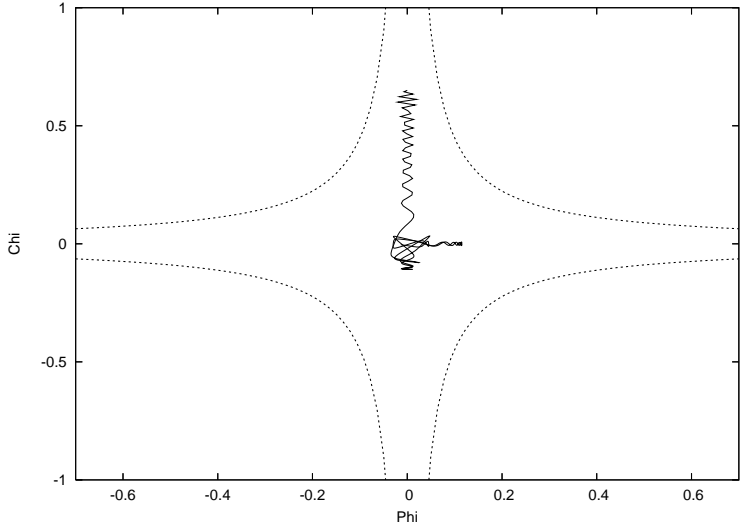,angle=-90,width=4in}
\caption{\label{500hubble} $\chi$ vs. $\phi$ until $t=500$ Planck times. Initial conditions here are $(\phi,\dot{\phi},\chi,\dot{\chi},a,\dot{a},b)=(0,-0.2,0.65,0,1,1.1,1)$ and $\dot{b}$ is determined by the Friedmann equation. The dotted line is the equipotential corresponding to the initial energy of the fields.}
\end{center}
\end{figure}
\begin{figure}[h]
\begin{center}
\epsfxsize=4.0in
\epsfig{figure=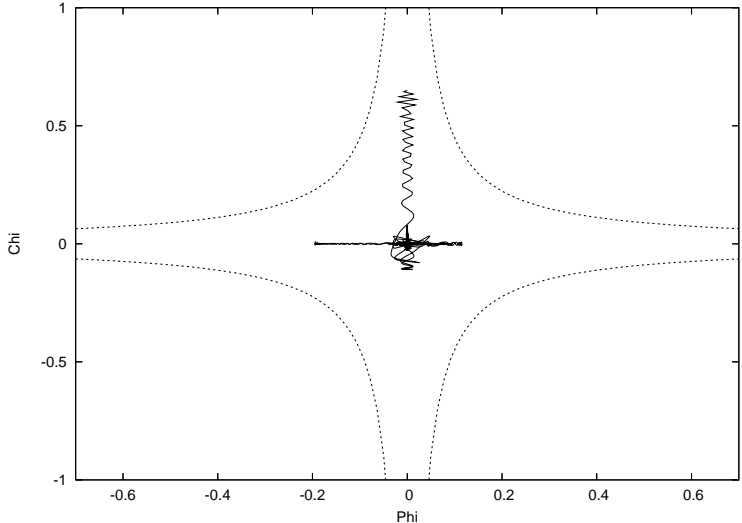,angle=-90,width=4in}
\caption{\label{15000hubble} $\chi$ vs. $\phi$ until $t=15000$ Planck times. Initial conditions are as for Figure \ref{500hubble}.}
\end{center}
\end{figure}

\begin{figure}[h]
\begin{center}
\epsfig{figure=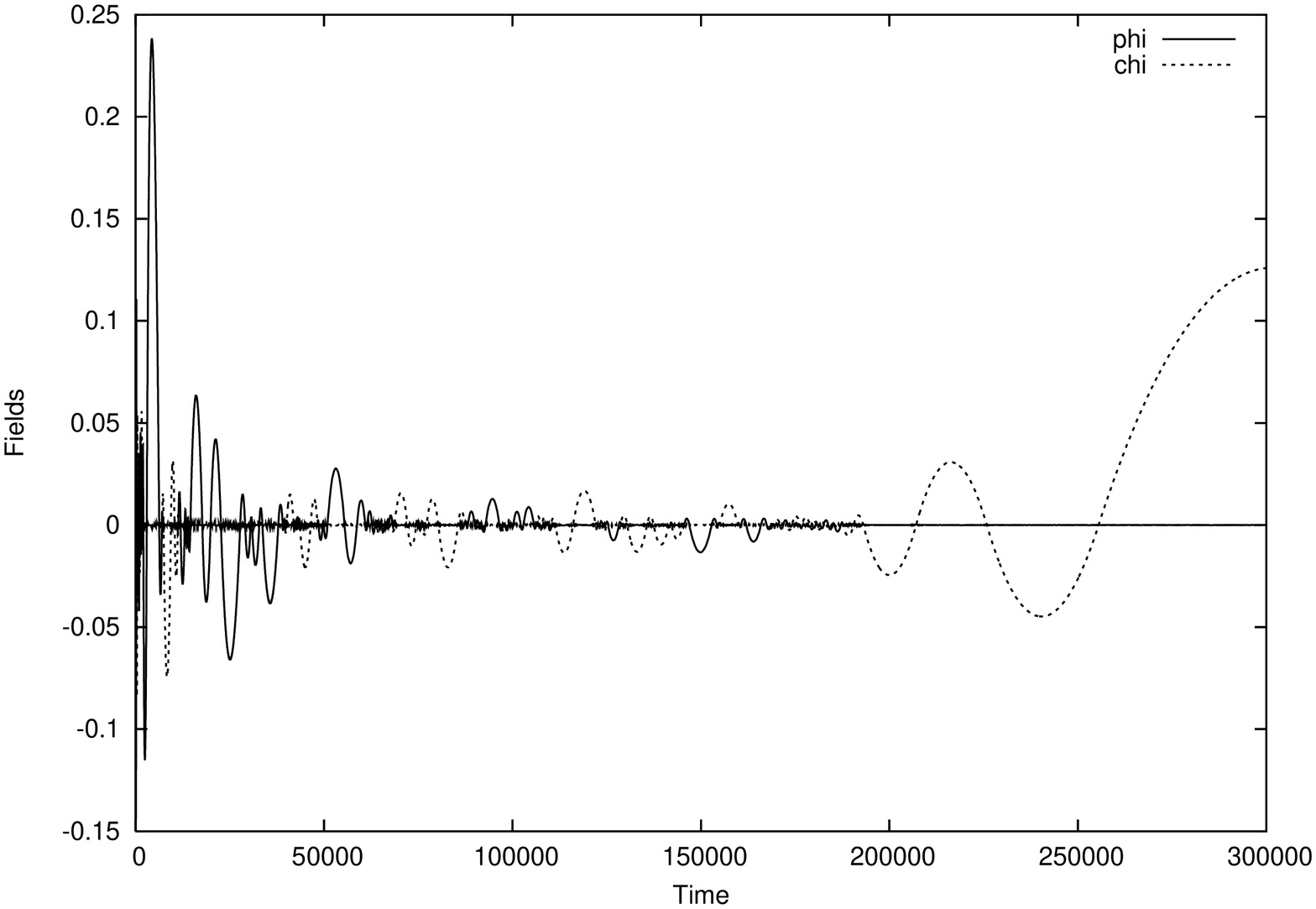,angle=-90, width=\textwidth}
\caption{\label{hnt} The fields $\phi$ and $\chi$ as functions of time. Looking only up to $t\sim 10^5$ one might think the fields were trapped at the origin. But then at $t\sim 2\times 10^5$ $\phi$ starts to increase. Initial conditions used here are: $(\phi,\dot{\phi},\chi,\dot{\chi},a,\dot{a},b)=(0,0.2,0.2,0,1,0.3,1)$ and $\dot{b}$ is determined by the Friedmann equation.}
\end{center}
\end{figure}

On the other hand, the effect of Hubble friction is to make the occasional sojourns down the arms rarer and longer lasting. By the time of nucleosynthesis ($t\sim 10^{45}$ Planck times), a typical journey down an arm may well last an extremely long time, so one might hope that over the time--scales we are interested in, the fields are essentially still. However, big bang nucleosynthesis (in particular the observed primordial Helium abundance) relies on a variety of processes that take place roughly between $t\sim 1$ sec, and $t\sim 3$ mins, and depend sensitively on the number of massless species of particle. The velocity of the moduli goes like $1/t$, so with an initial velocity $ v_0$, the distance in field space the moduli traverse during BBN is $\sim v_0\log(180)\sim 5v_0$ in Planck units. Since the mass of $\chi$ say is $g\phi$, we expect that over this time large changes to the masses of $\chi$ and $\phi$ are possible despite the slowing effect of Hubble friction. Put differently, for the change in mass to be less than a GeV, the initial velocity must be fine tuned to $v_0<10^{-19}$. The effect of Hubble friction is therefore still somewhat unpredictable.\footnote{Even if the fields were effectively still over the course of BBN,  there is no particular reason to believe that at the relevant time they should be in any particular place, i.e. near the origin, or down an arm.}

It is worth noting that simulations performed in \cite{Mohaupt:2004pr} up to $t\sim 1000 t_{\text{Planck}}$ suggest that the fields eventually get trapped near the origin. However, running the simulations for longer interval reveals that the apparent trapping at the origin can be misleading. Figure \ref{500hubble} shows the evolution of $\phi$ and $\chi$ up to $t=500t_{\text{Planck}}$, when it appears that they are becoming trapped. Figure \ref{15000hubble} shows the subsequent motion until $t=15000t_{\text{Planck}}$. During the intervening period the apparently stabilized field took a journey down the $\chi=0$ arm. Figure \ref{hnt} shows a different example of the same phenomenon, with the fields plotted against time. We can understand this as follows: with no Hubble friction, the fields are free to explore the full phase space between the lines of equipotential determined by the energy of the system. Once we turn Hubble friction on, the energy in the fields is no longer conserved and the available phase space decreases with time. However, because the arms are exactly flat it is always possible for the fields to be sent arbitrarily far down an arm. Since the scattering is chaotic this eventually happens. The result is that instead of slowing the chaotic motion to a halt, Hubble friction merely stretches the motion out onto longer and longer timescales.

These conclusions change significantly if one allows corrections to the $\phi^2\chi^2$ potential. If the corrections do not lift the vacua (i.e. the additional terms are of the form $\phi^m\chi^n$ with $m$ and $n$ both $\neq0$) then we still expect the fields to move down the arms every now and again, but there is no guarantee that they will return to the stadium. The derivation of the restoring force from the effective potential \eqref{effectivepot} depends crucially on the quadratic nature of $V(\phi,\chi)$ .

If, on the other hand, a vacuum branch is lifted, say by the addition of a $\phi^4$ term\footnote{This is precisely what one expects in the neighborhood of an arbitrary conifold point, i.e. one that is not necessarily a transition point between two topologically different families of Calabi--Yaus.}, then Hubble friction indeed does have a trapping effect. Reduction in energy gradually confines $\phi$ to a smaller and smaller region of field space, and as $t\to\infty$ the energy vanishes, so there is some definite value of $\phi$ that the field configuration approaches.

\section{Quantum Particle Production} \label{quantum}

Beyond the classical dynamics discussed in the previous section, essential quantum mechanical phenomena occur at ESPs. In particular, when a field is massless, on--shell excitations are produced at arbitrarily small energy cost.

Particle production at ESPs has been studied previously in the context of preheating models \cite{Traschen:1990sw,Kofman:1994rk,Kofman:1997yn,Felder:1998vq,Felder:1999pv}, and also in various string theory scenarios \cite{Lawrence:1995ct,Gubser:2003vk,Kofman:2004yc,Silverstein:2003hf,McAllister:2004gd,Watson:2004aq,Cremonini:2006sx,Friess:2004zk}. But it is of interest here because it results in a potential which confines the fields near the extra species point \cite{Kofman:2004yc}. The rough idea is that as one of the fields becomes massless, a burst of on--shell particles is produced. The mass of these particle increases if the fields subsequently move away from the ESP, so energetics force the fields to stay nearby.\footnote{Particle production is not the only quantum modification to the classical motion. There are also corrections to the kinetic terms and loop corrections to the potential (Coleman--Weinberg corrections). There is however a set of circumstances in which particle production is the most important effect. It was shown in \cite{Silverstein:2003hf} that for weak coupling ($g^2 \ll 1$) the effect of on-shell production dominates over propagator
corrections from virtual scattering.  This will be the case here as well, since the coupling $g$ must be small to ensure the validity of the SUGRA description.
The Coleman-Weinberg potential requires a bit more care, since after SUSY breaking loop corrections
should be expected to play an important role.  In \cite{Kofman:2004yc} it was argued that these corrections are subdominant because on-shell production always gives a positive contribution for both bosons and fermions, whereas fermions running in
loops will give an opposite contribution to the effective potential from that of bosons. If SUSY breaking is soft, this implies that the induced potential from on-shell production should
dominate the effective potential relative to the contribution from loops.  However, a better understanding of the effects of SUSY breaking remains an important goal for string cosmology.}

For example, consider the potential \eqref{minpot}, and for initial conditions take $\phi$ and $\chi$ approximately homogeneous in space, positioned along one of the vacuum branches, and moving towards the origin. For definiteness say $\chi=0$, $\phi>0$, and $\dot{\phi}<0$. The effective mass of the $\chi$ particles is $m_\chi=g\phi(t)$ which vanishes as $\phi$ crosses the origin. So a burst of on--shell $\chi$ particles is produced. Now $\phi$ can no longer move freely away from the origin since conservation of energy places a restriction on how massive the incoherent $\chi$ excitations can become. This results in a potential, trapping $\phi$ at the origin, which gets steeper as more $\chi$ quanta are produced. Since there is a $\phi^2\chi^2$ interaction term, scattering of $\chi$ particles results in excitations of the $\phi$ field as well, effectively driving both fields towards the origin.

The aim of this section is to study how particle production might change the conclusions of the classical analysis of the previous sections. We use the same metric ansatz as before:
\begin{align}
ds^2=-dt^2+a(t)^2 \ci{dx_1^2+dx_2^2+dx_3^2}+b(t)^2dx_4^2
\end{align}
with stress tensor $T^\mu_\nu=\text{diag}(-\rho, p,p,p, p)$. And for simplicity we take the fields to be moving along the $\chi=0$ arm initially. The equations of motion are then:
\begin{align}
    \ddot{\phi} + \ci{3H_a+H_b}\dot{\phi}  &= -  V'(\phi) \label{phionly1} \\
    \dot{H_a} + H_a\ci{3H_a+H_b} &= \frac{\kappa_5^2}{3}(\rho-p) \\
    \dot{H_b} + H_b\ci{3H_a+H_b} &= \frac{\kappa_5^2}{3}(\rho-p) \label{phionly3}
\end{align}
where $H_a = \dot{a}/a$ and $H_b = \dot{b}/b$ are the Hubble parameters. To study excitations of $\chi$, we substitute $\chi=\chi(t)+\delta\chi(t,\vec{x},y)$ into the $\chi$ equation of motion \eqref{chieqn}, which results in:
\begin{align}
\delta \ddot{\chi} +\left( 3H_a +H_b \right) \delta  \dot{\chi}- \left( \frac{1}{a(t)^2}\partial_{\vec{x}}^2 + \frac{1}{b(t)^2}\partial_{y}^2  -g^2\phi(t)^2 \right) \delta \chi=0,
\end{align}
A slightly modified Fourier transform gives a simple mode equation:
\begin{align}
    \delta \chi= &\int {\rm d}^3k\;{\rm d} k_4 \frac{\chi_k(t)}{\sqrt{a^3 b}}e^{i {k} \cdot \vec{x}+ik_4y} \\
 \label{chieom}
&\ddot{\chi}_k+\omega_k^2(t) \chi_k=0
\end{align}
where the frequency $\omega_k$ of the $k^{\text{th}}$ mode is:
\bea \label{omega0}
\omega_k^2(t)&=&\frac{\vec{k}^2}{a^2}+\frac{k_4^2}{b^2}+g^2 \phi^2(t)-\frac{1}{4}\left(  9H_a^2+H_b^2+6 H_a H_b+6 \dot{H}_x+2 \dot{H}_y  \right)\quad\quad.
\eea
To proceed we need to compute the contributions to the right hand sides of equations (\ref{phionly1}--\ref{phionly3}) (i.e. $V',\rho$ and $p$) resulting from a burst of particle production. This can be determined from $n_{\kappa}$, the number density of particles produced with \emph{physical} momentum $\kappa$, in the following way:
\begin{align}
    n_{\chi}& = \int\frac{d^4\kappa}{(2\pi)^4}n_{\kappa} = \text{number density in all modes} \\
    \rho_{\chi} &= \int\frac{d^4\kappa}{(2\pi)^4}\omega_{\kappa} n_{\kappa} \cm (\omega_{\kappa} = \omega_k \text{ with } (\vec{k},k_4)=(a\vec{\kappa},b\kappa_4))
\end{align}
To find the potential it is useful to recall the case with 2 fields: $V=\frac{1}{2}g^2\phi^2\chi^2$, and approximate $\chi^2\to \left<\chi^2\right>$. For the incoherent $\chi$ oscillations associated with production of $\chi$ particles, we have:
\begin{align}
    \an{\chi^2} = \frac{\rho_{\chi}}{g^2\phi^2} \simeq \frac{n_{\chi}}{g|\phi|}
\end{align}
since for slowly varying $\phi$, $\chi$ is approximately a harmonic oscillator with frequency $g\phi$.  In the last step we used $\rho_{\chi}\simeq g n_{\chi}|\phi|$, which amounts to approximating the frequency by the contribution to it from the effective mass only.\footnote{In other words, we assume $\omega_k^2\simeq g^2\phi^2 + f(k)$, and take $f(k)$ to be small. This is a good approximation because $f(k)$ is suppressed by powers of time, and we expect the time of production $t_{\#}$ to be $\gg t_P$ because natural initial field values are a distance $\sim 1$ from the origin, but initial velocities must be $\ll 1$ to justify an effective field theory description.} The derivative of the potential $V'(\phi)$ is then given by:
\begin{align}
    V'(\phi) = g^2\an{\chi^2}\phi = g n_{\chi} \frac{\phi}{|\phi|}
\end{align}
Finally, the pressure $p$ can be found by solving $\dot{\rho}=-(3H_a+H_b)(\rho+p)$, and using the $\phi$ equation of motion.

It is important to note that even if there is just a single production event, $n_{\chi}$ is a function of time, because the particles are diluted as the universe expands. Since  $n_k$ is appreciable only for $k\ll 1$ in the regime where effective field theory is useful, the produced excitations behave like dust:
\begin{align}
    n_{\chi}(t) = n_\chi(t_\#)\frac{a^3(t_\#)b(t_\#)}{a^3(t)b(t)}
\end{align}
where $t_{\#}$ is the time of production. The number density per mode $n_{\kappa}(t_{\#})$ was calculated for an isotropic background in \cite{Kofman:2004yc}, and the results generalize (as derived in the appendix) to our 5d case as follows:\footnote{In $D+1$ dimensions: $n_{\chi}\propto\int_0^{\infty} dk k^{D-1} \exp\ci{-\gamma \frac{k^2}{gv_{\#}}} \propto (gv_{\#})^{\frac{D}{2}}$, where $\gamma$ is a number of order 1}
\begin{eqnarray}
    n_{\chi}\ci{t_{\#}} &\propto& \ci{gv_{\#}}^2 \label{correctionterms1}  \\
    \rho &=& \frac{1}{2}\dot{\phi}^2 + \rho_{\chi} = \frac{1}{2}\dot{\phi}^2(t) + g n_\chi(t_\#)\frac{a^3(t_\#)b(t_\#)}{a^3(t)b(t)} |\phi (t)| \\ \label{correctionterms2}
    V'(\phi) &=& g n_{\chi}\ci{t_{\#}} \frac{a^3(t_{\#})b(t_{\#})}{a^3(t)b(t)} \frac{\phi(t)}{|\phi(t)|} \\ \label{correctionterms3}
    p &=& \frac{1}{2}\dot{\phi}^2
\label{correctionterms4}
\end{eqnarray}
where $v_{\#}=\frac{d\phi}{dt}|_{t=t_{\#}}$.

\subsection{Numerical Simulations and Comparison with Classical Effects}

Our aim now is to understand how production of on--shell excitations changes the results of section \ref{hubblefriction}. Rather than work with typical initial conditions, we consider the `worst case scenario' from the point of view of trapping, i.e. motion directly down one of the arms of the potential (classically there is no trapping at all: $\phi(t) \sim \log(t)$). Through this analysis, we intend to set a lower bound on the efficiency of trapping. With this in mind we include only a single particle production event, and do not alter the potential in subsequent crossings of the origin.

More general initial conditions do not allow the fields to move arbitrarily far from the origin. And it is argued in \cite{Kofman:2004yc} that successive particle production events are enhanced by parametric resonance since they take place in a bath of the particles already produced.

It is not easy to guess the net result of including particle production, since there are several competing effects:
\begin{itemize}
    \item Hubble friction slows down the motion of the fields
    \item The expanding background dilutes the produced particles thus reducing the slope of the induced potential.
    \item The particles backreact on the background
\end{itemize}

\begin{figure}
\begin{center}
\begin{picture}(0,0)%
\includegraphics{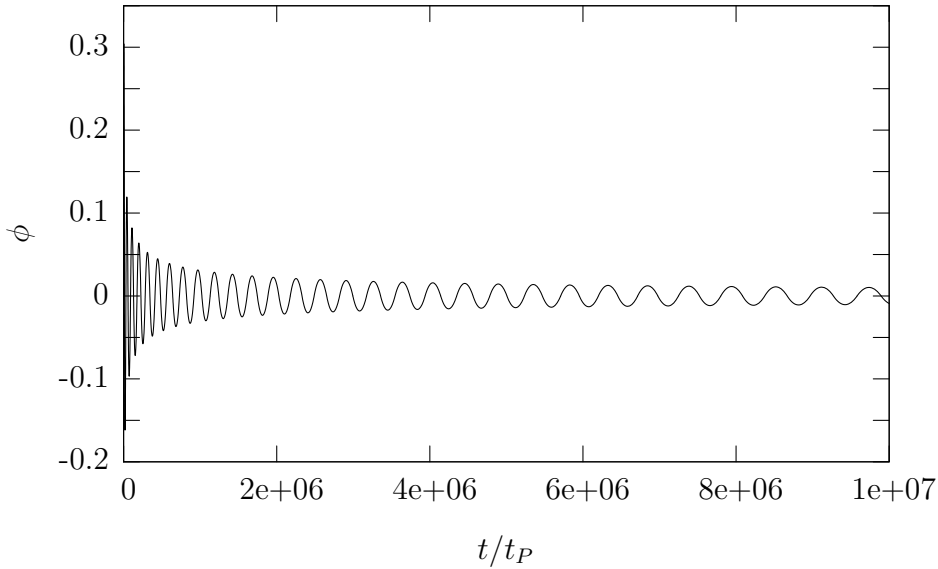}%
\end{picture}%
\begingroup
\setlength{\unitlength}{0.0200bp}%
\begin{picture}(18000,10800)(0,0)%
\put(2475,1650){\makebox(0,0)[r]{\strut{}-0.2}}%
\put(2475,3214){\makebox(0,0)[r]{\strut{}-0.1}}%
\put(2475,4777){\makebox(0,0)[r]{\strut{} 0}}%
\put(2475,6341){\makebox(0,0)[r]{\strut{} 0.1}}%
\put(2475,7905){\makebox(0,0)[r]{\strut{} 0.2}}%
\put(2475,9468){\makebox(0,0)[r]{\strut{} 0.3}}%
\put(2750,1100){\makebox(0,0){\strut{} 0}}%
\put(5635,1100){\makebox(0,0){\strut{} 2e+06}}%
\put(8520,1100){\makebox(0,0){\strut{} 4e+06}}%
\put(11405,1100){\makebox(0,0){\strut{} 6e+06}}%
\put(14290,1100){\makebox(0,0){\strut{} 8e+06}}%
\put(17175,1100){\makebox(0,0){\strut{} 1e+07}}%
\put(550,5950){\rotatebox{90}{\makebox(0,0)[t]{ $\phi$  }}}%
\put(9962,275){\makebox(0,0)[t]{ $t/t_{P}$ }}%
\end{picture}%
\endgroup
\caption{\label{sf1} Evolution of $\phi$ with time for typical initial conditions.}
\end{center}
\end{figure}

\begin{figure}
\begin{center}
\begin{picture}(0,0)%
\includegraphics{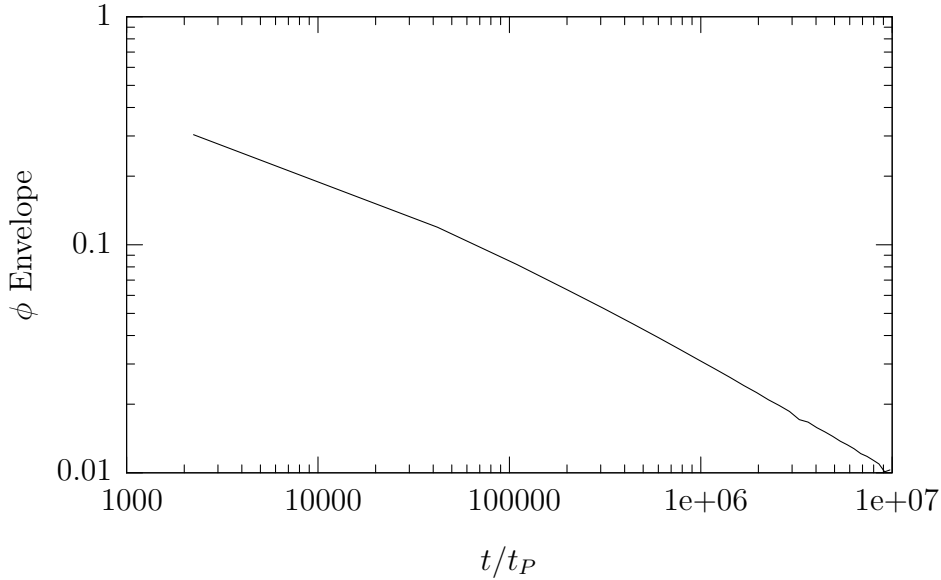}%
\end{picture}%
\begingroup
\setlength{\unitlength}{0.0200bp}%
\begin{picture}(18000,10800)(0,0)%
\put(2475,1650){\makebox(0,0)[r]{\strut{} 0.01}}%
\put(2475,5950){\makebox(0,0)[r]{\strut{} 0.1}}%
\put(2475,10250){\makebox(0,0)[r]{\strut{} 1}}%
\put(2750,1100){\makebox(0,0){\strut{} 1000}}%
\put(6356,1100){\makebox(0,0){\strut{} 10000}}%
\put(9963,1100){\makebox(0,0){\strut{} 100000}}%
\put(13569,1100){\makebox(0,0){\strut{} 1e+06}}%
\put(17175,1100){\makebox(0,0){\strut{} 1e+07}}%
\put(550,5950){\rotatebox{90}{\makebox(0,0)[t]{ $\phi$ Envelope }}}%
\put(9962,275){\makebox(0,0)[t]{ $t/t_{P}$ }}%
\end{picture}%
\endgroup
\caption{\label{sf2} Here the maxima of $\phi$ are plotted on a log--log scale, showing the power law decrease of the $\phi$ amplitude.}
\end{center}
\end{figure}

In \cite{Kofman:2004yc} the first two of these effects were taken into account, but the scale factor was assumed to be a power of time. Here we include all three effects. For typical initial conditions (with either a 4d or a 5d background), we find solutions like those in Figure \ref{sf1}. A plot of $\log(\phi)$ against $\log(t)$ (Figure \ref{sf2}) reveals a power law decrease in the amplitude of $\phi$ oscillations, suggesting that after $\phi$ is restricted to an ever--decreasing range of values, it approaches 0 like a negative power of $t$ as $t\to\infty$. Since this is a lower bound on trapping efficiency, we expect that motion with generic initial conditions (not along an arm of the potential) has the same features. This is to be contrasted with the classical situation of Figure \ref{hnt}, where the fields are not contained in an envelope with monotonically decreasing amplitude. Quantum particle production thus provides a viable mechanism for trapping some of the moduli at points where extra massless species appear.

\section{Conclusion}

We have considered the effects of additional light degrees of freedom on both the classical and semi-classical dynamics. In the purely classical case we have illustrated the appearance of chaotic scattering in the stadium region, and found that, somewhat surprisingly, vestiges of the chaotic dynamics are important even when Hubble resistance is taken into account. In particular, rather than slowing the fields to a halt, the effect of the expansion seems to be to stretch the motion out on ever larger timescales. The point is that the chaotic dynamics ensures that the weakest link in the stabilization mechanism will, sooner or later, be probed. And because the potential contains flat directions, these loci form the weak links and they allow fields to roll away from any would-be stabilization location. Moduli stablization therefore seems unlikely to result from classical motion in the quartic $\phi^2\chi^2$ potential. Higher order corrections may change this conclusion significantly, but the end result (i.e. where the fields finally rest) will still depend sensitively on initial conditions if the corrections do not lift the flat directions.

These caveats about stabilization do not apply once the effects of quantum particle production are taken into account. The amplitude of the expectation values of the fields decreases like a power of time. Though we only simulated the effect for the $\phi$ field, once incoherent $\chi$ excitations are present, the $\chi^2\phi^2$ term allows scattering to produce $\phi$ particles also, which produces a potential for $\chi$ similar to that for $\phi$.

An important problem that remains is to consider additional quantum effects.  In this semi-classical analysis we have only focused on on-shell particle production.
However, understanding propagator and loop corrections in the absence of supersymmetry is a vital next step.

One obstacle to using the above framework to stabilize moduli is that in the known models whose low energy theory includes a piece of the form \eqref{lag}, only a proper subset of the moduli possess interaction terms of the type required. For example in a Calabi--Yau compactification of M--theory, if we approach the conifold point by blowing down 2--spheres, then in general only the K\"{a}hler moduli will couple to the brane wrapping modes. Separate considerations, such as the inclusion of $p$-form flux windings or nonperturbative effects like gaugino condensation, are still needed to produce a phenomenologically acceptable theory.

One may also wonder whether the linear potential resulting from particle production really solves the phenomenological problems mentioned in the introduction, since it is not a mass term. However, the concept of particle number is only well defined far away from the time when $\phi$ crosses the origin, so it seems likely that the sharp cusp in the potential is smoothed out. A large mass term is then present, but large interactions also.

Finally, it is important to note that the potential arising from particle production is proportional to the number \emph{density} of the excitations. So as the universe expands, the trapping effect becomes weaker and weaker. Therefore, while it may help to ameliorate the issues moduli create with primordial nucleosynthesis, late--time effects such as varying couplings are still a problem.

The important case of particle production and dynamics in a contracting universe has not been previously studied and will be the subject of a future paper \cite{Weltman:2006}. It is not immediately obvious what the effect of contraction will be and it may be of great relevance to models of the universe that include a phase of contraction such as the cyclic universe \cite{cyclic}.

\vspace{2ex}
\textbf{Acknowledgments} 

We would like to thank Yashar Ahmadian, Dick Bond, Raphael Bousso, Sera Cremonini, Alex Hamilton, Bob Holdom, Lam Hui, Nemanja Kaloper, Lev Kofman, Andrei Linde, Jeff Murugan, Savdeep Sethi and Daniel Wesley for useful discussions.
BRG acknowledges financial support from DOE grant DE-FG-02-92ER40699.
The work of SJ is supported by the Pfister foundation.
JL and SJ acknowledge financial support from Columbia University ISE.
SW would like to thank the Galileo Galilei Institute for Theoretical Physics for hospitality and INFN for partial support during the completion of this work.
SW was also supported in part by the National Science and Engineering Research Council of Canada.
The work of AW is supported by a NASA Graduate Student Research Fellowship grant NNG05G024H.
AW would like to thank the participants of the session 86 Les Houches summer school for many useful conversations as well as CITA and the University of Cape Town for their hospitality during completion of this work.

\appendix

\section{Particle Production}
\begin{figure}
\begin{center}
\begin{picture}(0,0)%
\includegraphics{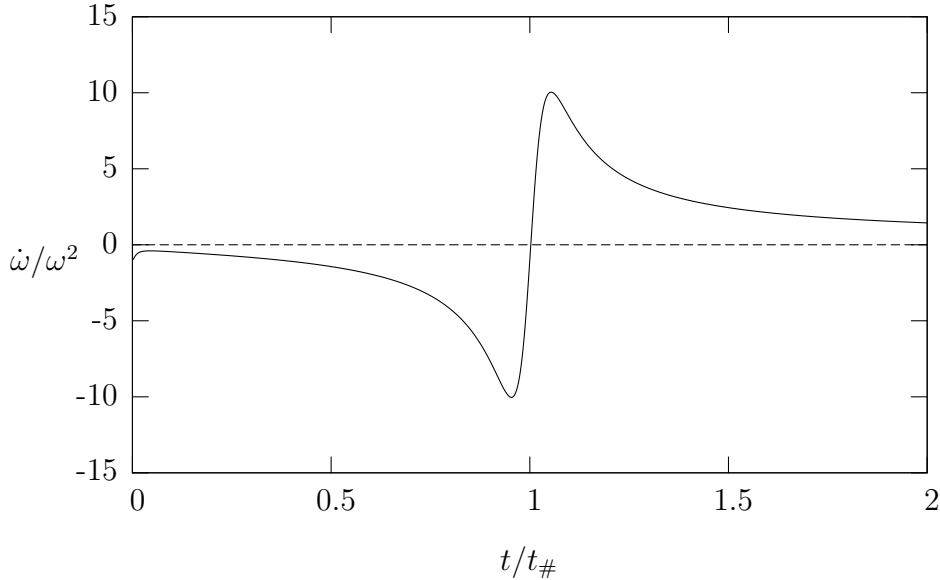}%
\end{picture}%
\begingroup
\setlength{\unitlength}{0.0200bp}%
\begin{picture}(18000,10800)(0,0)%
\put(1925,1650){\makebox(0,0)[r]{\strut{}-15}}%
\put(1925,3083){\makebox(0,0)[r]{\strut{}-10}}%
\put(1925,4517){\makebox(0,0)[r]{\strut{}-5}}%
\put(1925,5950){\makebox(0,0)[r]{\strut{} 0}}%
\put(1925,7383){\makebox(0,0)[r]{\strut{} 5}}%
\put(1925,8817){\makebox(0,0)[r]{\strut{} 10}}%
\put(1925,10250){\makebox(0,0)[r]{\strut{} 15}}%
\put(2200,1100){\makebox(0,0){\strut{} 0}}%
\put(5944,1100){\makebox(0,0){\strut{} 0.5}}%
\put(9688,1100){\makebox(0,0){\strut{} 1}}%
\put(13431,1100){\makebox(0,0){\strut{} 1.5}}%
\put(17175,1100){\makebox(0,0){\strut{} 2}}%
\put(550,5950){\rotatebox{0}{\makebox(0,0)[t]{ $\dot{\omega}/\omega^2 $ }}}%
\put(9687,275){\makebox(0,0)[t]{ $t/t_{\#}$ }}%
\end{picture}%
\endgroup
\caption{\label{nap} The non--adiabatic parameter $\dot{\omega}/\omega^{2}$ for very long wavelength modes (plotted here for $k=0$). The other parameters are $t_{\#}=10^5$ and $gv_{\#}=10^{-11}$, consistent with the effective field theory approximation. The sharpness of the peak increases with $t_{\#}$, and is independent of the choice of Kasner parameter $q$. }
\end{center}
\end{figure}

A formal solution to the mode equation (\ref{chieom}) is given by
\be
\chi_k(t) = \frac{\alpha_k(t)}{\sqrt{2\omega_k(t)}}e^{-i\int^t \omega_k(t')
dt'}+\frac{\beta_k(t)}{\sqrt{2\omega_k(t)}}e^{i\int^t \omega_k(t') dt'}.
\ee
Normalizing $\chi_k$ by using the Klein--Gordon inner product is then equivalent to
$|\alpha_k(t)|^2-|\beta_k(t)|^2=1$, which can be used to
write the equation of motion as
\bea \label{smeqns}
\dot{\alpha}_k &=& \frac{\dot{\omega}_k}{2\omega_k} e^{2i\int^t
\omega_k(t')dt'}\beta_k \nonumber \\ \dot{\beta}_k &=&
\frac{\dot{\omega}_k}{2\omega_k} e^{-2i\int^t
\omega_k(t')dt'}\alpha_k
\eea
If initially we have $\beta_k\ll 1$ and $\alpha_k \sim 1$ then at late times, the number density of particles produced in the $k^{\text{th}}$ mode is $n_k=|\beta_k|^2$. Assuming that $\beta_k$ is always small, the production can be estimated by:
\begin{align} \label{beta}
\beta_k \sim \int  dt
\frac{\dot{\omega}_k}{2\omega_k} e^{-2i\int^t
\omega_k(t')dt'}
\end{align}
We assumed earlier that the particle production happens in a short burst as $\phi$ crosses the origin. To justify this approximation, note that the integrand of \eqref{beta} is only significant when $\dot{\omega}_k \ll \omega_k$, or in terms of dimensionless quantities when:
\begin{align}
\frac{\dot{\omega_k}}{\omega_k^2} \gtrsim 1\quad\quad.
\end{align}
To get an idea of how $\frac{\dot{\omega_k}}{\omega_k^2}$ behaves, consider the following analytic solution of the equations of motion (\ref{phionly1}--\ref{phionly3}):
\begin{align}
    \phi(t) &= v_{\#}t_{\#}\log\ci{\frac{t}{t_{\#}}} \\
    H_a(t) &=\frac{q}{t} \\
    H_b(t) &=\frac{1-3q}{t}
\end{align}
which results in the following mode frequencies:
\begin{align} \label{modes}
\omega_k^2(t) = \frac{\vec{k}^2}{t^{2q}}+\frac{k_4^2}{t^{2(1-3q)}}+\frac{1}{4t^2} + g^2 v_{\#}^2t_{\#}^2\sq{\log\ci{\frac{t}{t_{\#}}}}^2
\end{align}
A natural initial field value is $\phi\sim 1$, while $v_\#\ll 1$ in order that an effective field theory description is appropriate. It follows that the time $t_{\#}$ when $\phi=0$ is large. Figure \ref{nap} shows a plot of $\dot{\omega}/\omega^2$ for typical values of $g$,$v_{\#}$ and $t_{\#}$, from which one can see that indeed a burst of particle production does occur at $t=t_{\#}$. Less obvious is that $\dot{\omega}/\omega^2$ is large near $t=0$ as well. This particle production is due to the expansion of the spacetime (essentially the $1/4t^2$ term in \eqref{modes}) rather than time dependence of $\phi$.
Here we will only calculate the effect of the latter, i.e. production at $t\simeq t_{\#}$.

For a general solution, the mode frequencies are:
\begin{align} \tag{\ref{omega0}}
\omega_k^2(t)= \frac{\vec{k}^2}{a^2}+\frac{k_4^2}{b^2}+g^2 \phi^2(t)-\frac{1}{4}\left(  9H_a^2+H_b^2+6 H_a H_b+6 \dot{H}_x+2 \dot{H}_y  \right)
\end{align}
and neglecting the gravitational terms (since $t_{\#}\gg t_{\text{Planck}}$) gives:
\begin{align}
\omega_k^2(t)\approx \frac{\vec{k}^2}{a(t)^2}+\frac{k_4^2}{b(t)^2}+g^2 \phi^2(t)
\end{align}

\begin{figure}
\begin{center}
\epsfxsize=9.5cm
\epsffile{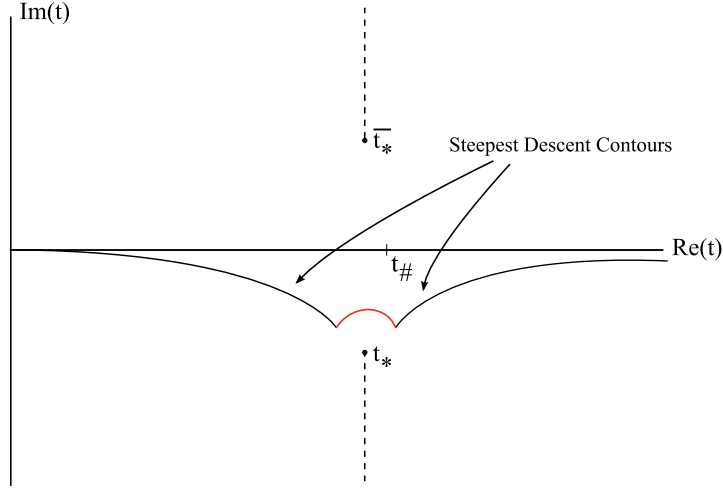}
\caption{\label{contour} The deformed contour for evaluating $\beta_k$. The arc near $t_*$ is the only part of the integral we evaluate.}
\end{center}
\end{figure}

We now estimate the integral \eqref{beta} following the method of \cite{Chung:1998bt}. Note first that the integral is dominated by times when $\omega$ is small. By allowing $t$ to take complex values, one can take advantage of the analytic structure of $\omega_k$ to compute the integral. In particular, one can deform the contour of integration so as to pass near a zero of $\omega_k$, located at a (necessarily complex) time $t_{\#}$ whose real part is close to $t_{\#}$. Since $\omega_k$ is real for real values of $t$, the zeros come in pairs related by complex conjugation, and from the single--valuedness of $\omega_k^2$, we see that $t_{\#}$ and $\bar{t_{\#}}$ are the branch points of a square--root branch cut, which for convenience we take to pass through $i\infty$ rather than along the real axis. The idea is that we will approximate $\beta_k$ by integrating around an small arc near $t_{\#}$ that interpolates between the two steepest descent paths, as in Figure \ref{contour}.

On this path we can expand $\omega_k$ in a Taylor series around $t_*$ whose terms are half--integral powers because of the branch point. In general, if $f(x)$ is analytic in a neighborhood of $x=0$ and $f(0)=0$, then we have:
\begin{align}
  \sqrt{f(x)} = \frac{1}{\sqrt{f'(0)}}\Bsq{f'(0)x^{1/2} + \frac{1}{4}f''(0)x^{3/2} + \frac{1}{4}\ci{\frac{1}{3}f'''(0)-\frac{1}{8}\frac{f''(0)^2}{f'(0)}}x^{5/2}+ \ldots }
\end{align}
The frequency therefore becomes:
\begin{align} \nonumber
    \omega_k (t) &= \sq{\sqrt{-2\frac{k^2}{a(t_*)^2}H_a(t_*) - 2\frac{k_4^2}{b(t_*)^2}H_b(t_*) +2g^2\phi(t_*)\dot{\phi}(t_*) } }(t-t_*)^{1/2} + \cO(t-t_*)^{3/2}\\ \label{freqtaylor}
    &= \sqrt{f(t_*)}(t-t_*)^{1/2} + \cO(t-t_*)^{3/2}
\end{align}
and integrating with respect to time:
\begin{align}
    \int_{0}^t \omega_k (t') dt' = \int_{0}^{t_*} \omega_k(t')dt' + \frac{2}{3}\sqrt{f(t_*)} (t-t_*)^{3/2} + \ldots
\end{align}
Substituting the above expressions into equation \eqref{beta}, one finds:
\begin{align}
\beta_k = e^{-2i\int_{0}^{t_*} \omega_k(t')dt'}\ci{ \frac{1}{4}\int_{C}\frac{d\delta}{\delta}e^{\frac{-4i}{3}
\sqrt{f(t_*)}\delta^{3/2}} }
\end{align}
where $\delta = t-t_*$ the $C$ denotes the red arc in Figure \ref{contour}   interpolating between the two steepest descent contours. The bracketed integral can be performed by a change of variables $\mu=\frac{4}{3}i\sqrt{f(t_*)}\delta^{3/2}$ which (see \cite{Chung:1998bt}) closes the contour $C$ to a loop as well as simplifying the exponential. The result is:
\begin{align}
\beta_k = \frac{i\pi}{3} \exp\ci{-2i\int_{-\infty}^{t_{\#}} \omega_k(t')dt'} \,
\exp\ci{-2i\int_{t_{\#}}^{t_*} \omega_k(t')dt'}
\end{align}
The first integral is real, contributing only a phase to the exponential which will cancel in $|\beta_k|^2$. The second integral can be approximated
$\int_{t_{\#}}^{t_*} \omega_k(t')dt' \sim \frac{i\gamma}{2} \text{Im} (t_*) \omega_k(t_{\#})$, where $\gamma$ is a constant of order 1. The result is:
\begin{align} \label{betaform}
|\beta_k|^2 = \biggl(\frac{\pi}{3}\biggr)^2 e^{\gamma\text{Im}(t_*)\omega_k(t_{\#})}.
\end{align}
Thus, to calculate the leading contribution to particle production
it is sufficient to identify the real and imaginary contributions
to the zeros of $\omega_k$.\footnote{We only get a sensible answer by deforming the contour into the lower half plane. This can be shown \cite{Chung:1998bt} to be a consequence of choosing the positive square root in \eqref{freqtaylor}.} For small $k$, we expect $t_*-t_{\#}$ to be small, so the following approximation is useful:
\begin{align} \label{freqsquaretaylor}
    0 = \omega_k^2(t_*) \simeq c_0 + c_1(t_*-t_{\#}) + c_2(t_*-t_{\#})^2
\end{align}
where the Taylor coefficients are:
\begin{align}
c_0&=\frac{k^2}{a^2(t_{\#})}+\frac{k_4^2}{b^2(t_{\#})}  \\
c_1&= -2\frac{k^2}{a^2(t_{\#})}H_a(t_{\#})-2\frac{k_4^2}{b^2(t_{\#})}H_b(t_{\#})  \\
c_2&= \cO(k^2) + g^2\dot{\phi}^2(t_{\#})
\end{align}
Here we have used $\phi(t_{\#})=0$. We can now find the zeros of $\omega_k$ by solving \eqref{freqsquaretaylor}. As argued in more detail in \cite{Chung:1998bt}, we are interested in the zero in the lower half plane:
\begin{align}
t_*=\frac{-c_1 - \sqrt{c_1^2-4 c_0 c_2} }{2c_2}
\end{align}
For small $k$ (so that terms $\cO(k^4)$ can be neglected), the comoving number density per mode is:
\begin{align}
    n_k(t_{\#}) &\simeq \ci{\frac{\pi}{3}}^2 e^{\gamma\text{Im}(t_*)\omega_k(t_{\#})} \simeq \ci{\frac{\pi}{3}}^2 e^{- \gamma c_0 /\sqrt{c_2}}\\ &\simeq \ci{\frac{\pi}{3}}^2  \exp\ci{ - \gamma \frac{1}{g v_{\#}} \sq{\frac{k^2}{a^2(t_{\#})}+\frac{k_4^2}{b^2(t_{\#})}}}
\end{align}
The total number density (in physical rather than comoving coordinates) in all modes is then found by integrating:
\begin{align}
    n_{\chi}(t_{\#}) &= \frac{1}{a^3(t_{\#})b(t_{\#})}\ci{\frac{\pi}{3}}^2 \int \frac{d^4 k}{(2\pi)^4} \exp\ci{ - \gamma \frac{1}{g v_{\#}} \sq{\frac{k^2}{a^2(t_{\#})}+\frac{k_4^2}{b^2(t_{\#})}}} \\ &= \frac{g^2v_{\#}^2}{144\gamma^2}\quad\quad.
\end{align}

\end{spacing}

\end{document}